\documentclass[twocolumn,english,floatfix,superscriptaddress,prl,showpacs]{revtex4-1}
\usepackage[T1]{fontenc}
\usepackage[utf8]{inputenc}
\usepackage{color}
\usepackage{amsmath}
\usepackage{graphicx}

\makeatletter

\@ifundefined{textcolor}{}
{%
 \definecolor{BLACK}{gray}{0}
 \definecolor{WHITE}{gray}{1}
 \definecolor{RED}{rgb}{1,0,0}
 \definecolor{GREEN}{rgb}{0,1,0}
 \definecolor{BLUE}{rgb}{0,0,1}
 \definecolor{CYAN}{cmyk}{1,0,0,0}
 \definecolor{MAGENTA}{cmyk}{0,1,0,0}
 \definecolor{YELLOW}{cmyk}{0,0,1,0}
}

\renewcommand\[{\begin{equation}}
\renewcommand\]{\end{equation}}

\makeatother

\usepackage{babel}
\begin{document}

\title{Non-gaussian spatial correlations dramatically weaken localization}

\author{H. Javan Mard}

\affiliation{Department of Physics and National High Magnetic Field Laboratory,
Florida State University, Tallahassee, FL 32306}

\author{E. C. Andrade}

\affiliation{Institut für Theoretische Physik, Technische Universität Dresden,
01062 Dresden, Germany}

\author{E. Miranda}

\affiliation{Instituto de Física Gleb Wataghin, Unicamp, R. Sérgio Buarque de
Holanda, 777, Campinas, SP 13083-859, Brazil}

\author{V. Dobrosavljević}

\affiliation{Department of Physics and National High Magnetic Field Laboratory,
Florida State University, Tallahassee, FL 32306}
\begin{abstract}
We perform variational studies of the interaction-localization problem
to describe the interaction-induced renormalizations of the effective
(screened) random potential seen by quasiparticles. Here we present
results of careful finite-size scaling studies for the conductance
of disordered Hubbard chains at half-filling and zero temperature.
While our results indicate that quasiparticle wave functions remain
exponentially localized even in the presence of moderate to strong
repulsive interactions, we show that interactions produce a strong
decrease of the characteristic conductance scale $g^{*}$ signaling
the crossover to strong localization. This effect, which cannot be
captured by a simple renormalization of the disorder strength, instead
reflects a peculiar non-Gaussian form of the spatial correlations
of the screened disordered potential, a hitherto neglected mechanism
to dramatically reduce the impact of Anderson localization (interference)
effects. 
\end{abstract}

\date{\today}


\maketitle
According to the scaling theory of localization \cite{Abrahams1979},
any amount of disorder suffices to localize all (non-interacting)
electrons at $T=0$ in dimension $d\le2$. In the presence of electron-electron
interactions, however, no such general statement exists, and the transport
behavior of disordered interacting electrons remains an outstanding
open problem \cite{dobrosavljevic2012conductor}. Since the relevant
analytical results are available only in some limiting cases \cite{efros1975coulomb,Lee-Ramakrishnan1985rmp,punnoose_sci2005},
complementary computational methods play a crucial role in providing
insight and information. Several numerical approaches have been recently
utilized to investigate transport properties of these systems, including
variational Hartree-Fock (HF) \cite{Herbut2001,heidarian04,wortis08}
and slave boson (Gutzwiller approximation) \cite{tanaskovi?2003disorder}
methods, as well as (numerically exact) quantum Monte Carlo techniques
\cite{denteneer99,Shepelyansky2003prb,fleury08a}. 

These studies provided evidence that repulsive electron-electron interactions
generally increase the conductance in small systems, with the suppression
of electronic localization being tracked down to partial screening
of the disorder potential. In principle, interactions could modify
either the amplitude or the form of spatial correlations \cite{Herbut2001}
of the renormalized disorder potential. The former mechanism is known
to be significantly enhanced by strong correlation effects \cite{tanaskovi?2003disorder}
and to survive even in high dimensions, while the latter is more pronounced
\cite{eric10} in the weak-coupling regime and in low dimensions \cite{Lee-Ramakrishnan1985rmp}. 

Despite this progress, several important questions remained unanswered:
(1) What is the dominant physical mechanism for disorder screening,
and can it \emph{qualitatively} modify the noninteracting picture?
(2) Can the interaction effects overcome Anderson localization and
stabilize the metallic phase in low dimensions? The task to carefully
and precisely answer these important questions in a model calculation
is the the main goal of this Letter. To do this, we utilize two different
variational methods to describe the statistics of the renormalized
disorder potential in an idealized dirty Fermi liquid\textcolor{black}{.
In contrast to most previous attempts, here we perform a careful finite
size scaling analysis of the conductance, which allows us to reach
conclusive results for the transport properties of the model we consider. }

\emph{Model and method.---} We study\textcolor{black}{{} the paramagnetic
phase of }a disordered Hubbard model 

\begin{equation}
H=-t\sum_{i,j,\sigma}(c_{i\sigma}^{\dagger}c_{j\sigma}^{\phantom{\dagger}}+\mathrm{h.c.})+\sum_{i,\sigma}\varepsilon_{i}n_{i\sigma}+U\sum_{i}n_{i\uparrow}n_{i\downarrow},\label{eq:ham}
\end{equation}
where $t$ is the hopping amplitude between nearest-neighbor sites,
$c_{i\sigma}^{\dagger}$$\left(c_{i\sigma}^{\phantom{\dagger}}\right)$
are the creation (annihilation) operators of an electron with spin
$\sigma=$ at site $i$, $U$ is the on-site Hubbard repulsion, and
$n_{i\sigma}=c_{i\sigma}^{\dagger}c_{i\sigma}^{\phantom{\dagger}}$.
The\emph{ spatially uncorrelated} random site energies $\varepsilon_{i}$
are drawn from a uniform distribution of zero mean and width $W$.
We work at half-filling, in units such that $t=a=e^{2}/h=1$, where
$a$ is the lattice spacing, $h$ is Planck's constant, and $e$ is
the electron charge. To be able to carry out the large scale computations
needed for conclusive finite-size scaling of the conductance, we focus
our attention on a one-dimensional model. Within the variational description
of a dirty Fermi liquid we consider, we expect the main trends to
persist in higher dimensions. 

Our starting point is the non-magnetic HF scheme \cite{Herbut2001},
where the renormalized site energies $v_{i}$ are given by
\begin{equation}
v_{i}=\varepsilon_{i}+\frac{U}{2}\left\langle n_{i}\right\rangle .\label{eq:renormenergies}
\end{equation}
\textcolor{black}{Here, $\left\langle n_{i}\right\rangle =\sum_{\sigma}\left\langle n_{i\sigma}\right\rangle $,
the average site occupation, is determined self-consistently in the
ground state, for each disorder realization. To cross-check our HF
predictions within a theory that is able to capture strong correlation
effects, we repeated the same calculations using the slave boson (SB)
mean-field theory (i.e. the Gutzwiller approximation) of Kotliar and
Ruckenstein \cite{kotliar_ruckenstein,tanaskovi?2003disorder}, generalized
to disordered systems} \cite{eric10}\textcolor{black}{. The SB theory
features two local variational parameters: the renormalized site energies
$v_{i}$ and the quasiparticle weight $Z_{i}$ ($Z_{i}=1$ within
HF) \cite{EGP2d_09,eric2009,suppl}. We found that, for moderate interaction
strength (not close to the Mott transition) and the low-dimensional
situation we consider, both methods produce qualitatively the same
behavior (see Fig.~\ref{fig:fullconductscaling} below), dominated
by a peculiar type of spatial correlation of the screened disorder
potential. The strong correlation effects (corresponding to $Z_{i}\ll1)$
do not appear to play a significant role in this regime (in contrast
to the situation explored in Ref. \cite{eric2009,eric10}). This makes
it possible to search for the relevant screening mechanism within
the simpler and physically very transparent HF scheme, which we focus
on in presenting most of our results. }

\textcolor{black}{To study the nature of the ground state we focus
on the dimensionless conductance $g$, which we obtain applying the
standard Landauer approach to our quasiparticle Hamiltonian \cite{suppl,eric12,Box-Muller,haugjauhobook}.
We numerically calculate $g$ in a set-up where we attach our system
to two non-interacting metallic leads at its ends \cite{nikolic01}.
For simplicity, we consider the wide band limit, where the leads'
self-energies are simply given by $\Sigma_{1\left(L\right)}=-i\eta/2$
\cite{suppl}, and in all our results we consider $\eta=1.0t$ (we
carefully checked that all our conclusions are independent of $\eta$
\cite{suppl}). Since we are working in one dimension, the conductance
displays wide sample to sample fluctuations}. We therefore focus on
its\emph{ typical} value, as given by the geometrical average $g=g_{typ}=\exp\overline{\ln g_{s}}$
\textcolor{black}{\cite{Anderson1980,markos2006numerical}}. In every
case, we averaged our results over $2,000$ disorder realizations,
which was sufficient to obtain very accurate results. 

\begin{figure}[t]
\begin{centering}
\includegraphics[width=3.5in]{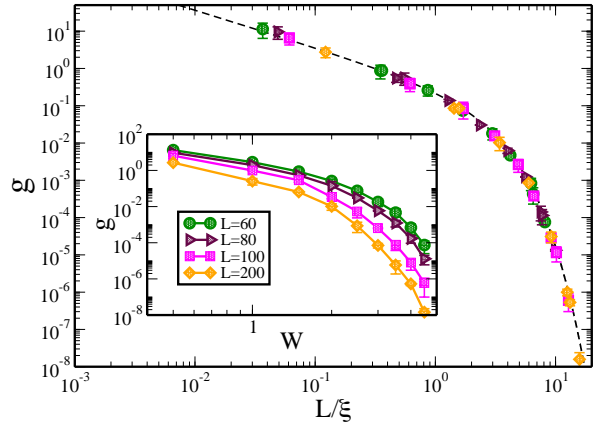}
\par\end{centering}

\protect\caption{\label{fig:gofL} The conductance scaling function for $U=1.0t$ obtained
with system sizes $L=60,\ 80,\ 100,$ and $200$ at several disorder
levels $W\le4.5t$. The dashed line is a modified version of the one-dimensional
conductance scaling function proposed in ref. \cite{Anderson1980}
(see also Eqs. \eqref{eq:fullscalingfunction} and \eqref{eq:intbetaand}):
$g=g^{*}/(\exp(x)-1)$, with $x=L/\xi$ and $g^{*}=0.366$. In the
inset, the conductance as a function of the disorder level is shown
for fixed system sizes.}
\end{figure}

\emph{Conductance scaling.---} In the non-interacting limit, the dependence
of the conductance on disorder and system size can be expressed in
a simple scaling function $g_{0}\left(x\right)$, with $x=L/\xi$,
where $\xi=\xi\left(W\right)$ is the localization length \cite{Abrahams1979,mackinnon81}.
Specifically, $g_{0}\propto L^{d-1}$ for $g\gg g^{*}$ (ohmic regime,
$x\ll1$) and $g_{0}\propto\mbox{exp}\left(-L/\xi\right)$\textcolor{magenta}{{}
}for $g\ll g^{*}$ (localized regime, $x\gg1$), where $g^{*}$ is
the characteristic dimensionless conductance which marks the crossover
between these two regimes. In particular, we use the expected exponential
decay of the conductance to determine $\xi$ for fixed values of $W$. 

\textcolor{black}{Using this scaling Ansatz, }we can collapse the
system size dependence of the conductance onto a scaling curve $g\left(L/\xi\right)$
\textcolor{black}{even in the presence of interactions,} as shown
in Fig.~\ref{fig:gofL}. Note that the error bars are approximately
the size of the data symbols or even smaller. We find that the localization
length increases considerably with $U$ (see \cite{suppl} for more
details). This enhancement of the localization length with interactions
has been often observed in studies of disordered interacting systems
\cite{heidarian04,henseler2008static,fleury08a,fleury08b}. We should
nevertheless stress that, despite the huge enhancement of $\xi$ with
$U$, there is always an exponential decrease for large $L$ and we
do not see any evidence of extended states.

Interestingly, all the curves $g\left(L/\xi\right)$ for different
interaction strengths can be made to collapse onto a single universal
curve by a proper interaction-dependent rescaling of the conductance,
see Fig.~\ref{fig:fullconductscaling}%
\footnote{Since the error bars are roughly the size of the symbols, we omit
them from now on for clarity.%
}. We call the conductance rescaling factor $g^{*}(U)$, and stress
that $g^{*}$ is a function of $U$ only. Its $U$-dependence for
both HF and SB approaches is shown in the inset of Fig.~\ref{fig:fullconductscaling},
where an exponential decrease with $U$ fits well the data in both
cases. 

\begin{figure}[t]
\begin{centering}
\includegraphics[width=3.2in]{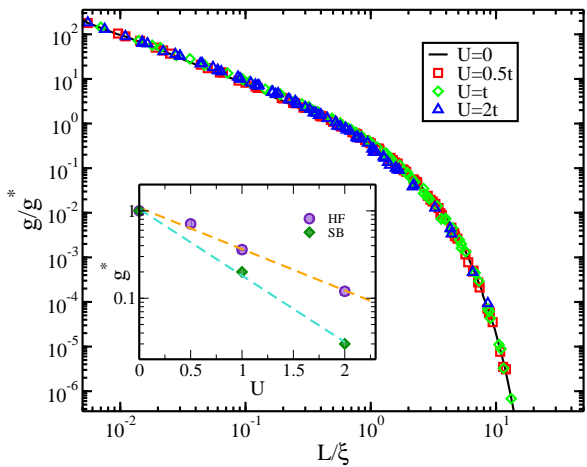}
\par\end{centering}

\protect\caption{\label{fig:fullconductscaling} Conductance scaling curves for different
$U$ values collapse onto a single universal curve after rescaling
both $L$ (by $\xi$) and $g$ {[}by $g^{\ast}\left(U\right)${]}.
In the inset, the characteristic conductance $g^{\ast}$ is plotted
as a function of $U$ for both HF and SB approaches. The behavior
is well fitted by an exponential: $g^{*}\left(U\right)=1.09(8)\exp\left[-1.09(7)U\right]$
(HF) and $g^{*}\left(U\right)=1.05\left(2\right)\exp\left[-1.75\left(3\right)U\right]$
(SB). }
\end{figure}

The above scaling implies that the full disorder and interaction dependence
of the conductance can be written as 
\begin{equation}
g=g^{\ast}\left(U\right)g_{0}\left[L/\xi\left(W,U\right)\right],\label{eq:fullscalingfunction}
\end{equation}
where $g_{0}\left(x\right)$ is the \emph{non-interacting scaling
function}, $g^{*}\left(U\right)$ sets the crossover conductance which
separates the weak localization regime ($g\gg1$) from the strongly
localized one ($g\ll1$), and we have explicitly shown all the $W$
and $U$ dependences.

\begin{figure}[t]
\begin{centering}
\includegraphics[scale=0.3]{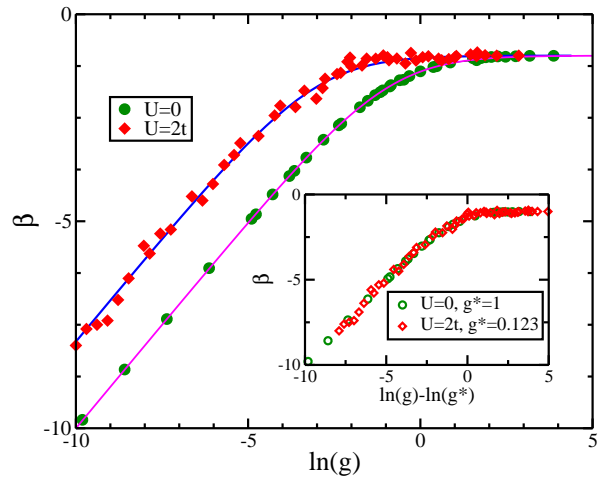}
\par\end{centering}

\protect\caption{\label{fig:betafunction} The beta function for $U=0$ and $U=2t$.
The numerical data are well described by Eq.~\eqref{eq:intbetaand}
with $g^{*}\left(U=0\right)=1.0$ and $g^{*}\left(U=2t\right)=0.123$.
Inset: The curve for $U=2t$ collapses onto the non-interacting one
with a shift of $\ln(g^{*})$ along the horizontal axis.}

\end{figure}

The scaling function in Eq.~\eqref{eq:fullscalingfunction} can then
be used to generate the beta function $\beta\left(g\right)=d\ln g/d\ln L$.
It follows immediately that the only effect introduced by interactions
on $\beta\left(g\right)$, as compared to its non-interacting counterpart,
is the rescaling of $g$ by the characteristic conductance $g^{\ast}\left(U\right)$
\begin{equation}
\beta\left(g\right)=\beta_{0}\left[g/g^{*}\left(U\right)\right],\label{eq:intbeta}
\end{equation}
where $\beta_{0}\left(g\right)$ is non-interacting beta function.
In particular, if we use the form of $\beta_{0}\left(g\right)$ proposed
in ref.~\cite{Anderson1980} we obtain

\begin{equation}
\beta\left(g\right)=-\left[1+\frac{g}{g^{*}\left(U\right)}\right]\mbox{ln}\left[1+\frac{g^{*}\left(U\right)}{g}\right].\label{eq:intbetaand}
\end{equation}
The validity of Eq.~\eqref{eq:intbetaand} can be double-checked
through a direct examination of the behavior of the beta function
for different values of $U$, as shown in Fig.~\ref{fig:betafunction}.
We stress that interaction-induced renormalizations of the localization
length alone are not capable of describing the results of Fig.~\ref{fig:betafunction},
as they \emph{drop out} of the beta function. Finally, using \eqref{eq:intbetaand},
we are able to give an operational definition of the characteristic
conductance: $g=g^{\ast}\left(U\right)$ at $L=L^{*}=(\ln2)\xi\left(W,U\right)$
\cite{suppl}.

\begin{figure}[t]
\begin{centering}
\includegraphics[scale=0.3]{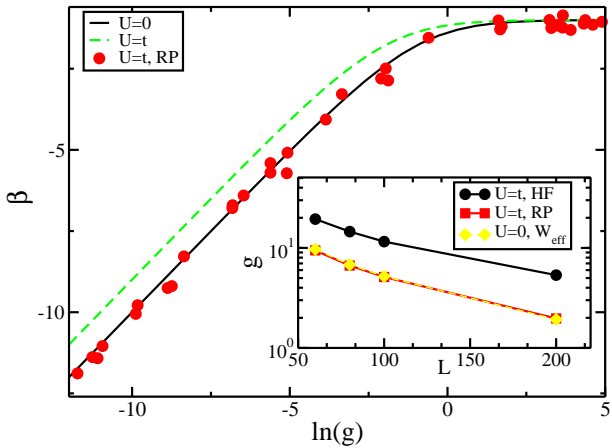}
\par\end{centering}

\protect\caption{\label{fig:randomizedbeta}Beta function for $U=1t$ obtained after
randomizing the self-consistently determined HF renormalized energies
$v_{i}$'s (red circles; see also the text for an explanation of the
procedure). Black and green lines are plotted using Eq.~\eqref{eq:intbetaand}.
In the inset, the conductance $g$ is shown as a function of $L$
for: (a) a system with $W=0.5t$ and $U=1t$ in the HF approximation
(black dots), (b) a non-interacting system whose site energies are
a random permutation of the renormalized site energies $v_{i}$ of
the HF approximation (red squares), and (c) a non-interacting system
system with $U=0$ and $W=W_{eff}=0.41t$ (gold diamonds).}
\end{figure}

\emph{Disorder screening and non-Gaussian spatial correlations.---}
A commonly invoked explanation for this conductance enhancement is
the fact that interactions act to ``screen'' the one-body potential
\cite{Herbut2001,tanaskovi?2003disorder,eric2009}. Within a mean-field
picture, an electron moving in the one-body potential $v_{i}$ ``sees''
site energies renormalized by the average interaction with the other
electrons, as in Eq.~\eqref{eq:renormenergies}. In the inset of
Fig.~\ref{fig:randomizedbeta}, we compare the conductance in the
full HF calculation for $W=0.5t$ and $U=1t$ with the one obtained
in the non-interacting case with an effective disorder $W_{eff}$
obtained from the width of the $v_{i}$ distribution \cite{suppl}.
It is clear that the screening effect by itself is not enough to reproduce
the conductance enhancement of the full HF calculation. This is further
confirmed when\textcolor{black}{, }after obtaining the fully converged
self-consistent HF values of $v_{i}$'s, we then calculate the conductance
of a \emph{non-interacting system} whose site energies are a random
permutation (RP) of the same $v_{i}$'s. Not surprisingly, the conductance
of the randomized system is essentially the same as the one for the
non-interacting system with \emph{uncorrelated} site energies distributed
uniformly with strength $W_{eff}$ (inset of Fig.~\ref{fig:randomizedbeta}).
In the main panel of Fig.~\ref{fig:randomizedbeta}, we also show
the beta function obtained from the RP of the HF results. As can be
seen, it reduces to the non-interacting one.\textcolor{black}{{} }The
effect of a RP of the renormalized site energies is to eliminate the
\emph{spatial correlations} between them. In the following we argue
that it is precisely these correlations which shift the crossover
scale $g^{*}\left(U\right)$ to much smaller values as compared to
the $U=0$ case. 

\begin{figure}[t]
\begin{centering}
\includegraphics[scale=0.3]{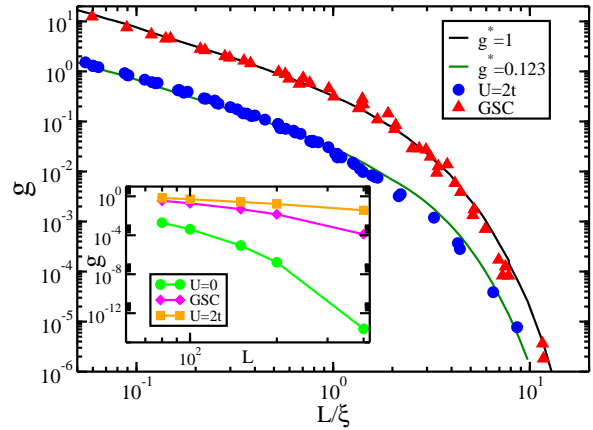}
\par\end{centering}

\centering{}\protect\caption{\label{fig:gaussscaling} Comparison of the conductance scaling function
of a disordered chain with $U=2t$ in the HF approximation and the
corresponding non-interacting system with the same gaussian correlated
site energies (GSC) (the inset shows the unscaled data). Solid lines
are drawn by using $g=g^{*}/(\exp(x)-1)$, as in Fig. \eqref{fig:gofL}.}
\end{figure}

\textcolor{black}{To further elucidate the pivotal role of spatial
correlations, we start by looking at the limit of weak disorder $W\rightarrow0$.
A perturbative calculation shows that the correlations among the $v_{i}$'s
are given by $\left\langle v_{i}v_{j}\right\rangle \sim r_{ij}^{-1}$,
for $r_{ij}\gg1$, where $r_{ij}=\left|r_{i}-r_{j}\right|$ \cite{suppl,eric10}.
These long-ranged correlations of the effective disorder potential
come from the usual Friedel oscillations. When properly tailored,
a correlated disorder potential may drive a metal-insulator transition
in $d=1$ \cite{dunlap90,moura98,IzrailevPRL82,garcia09,petersen13}.
In order to go beyond weak disorder, we first generate numerically
the two point correlation function $\left\langle v_{i}v_{j}\right\rangle $
from our HF results \cite{suppl}. We then implement a standard procedure
to generate random $v_{i}$'s with }\textcolor{black}{\emph{gaussian
correlations}}\textcolor{black}{{} of zero mean and covariance matrix
$\left\langle v_{i}v_{j}\right\rangle $ (note that the generated
data have no correlations beyond gaussian). Finally, we calculate
the conductance of a non-interacting system with the latter site energies.
Essentially, we want to know if the gaussian correlations contained
in $\left\langle v_{i}v_{j}\right\rangle $ are sufficient to account
for the $g^{\ast}$ renormalization. Fig.~\ref{fig:gaussscaling}
displays the results of this numerical procedure (which we dubbed
gaussian spatial correlations (GSC)).}\textcolor{magenta}{{} }\textcolor{black}{Although
the conductance is enhanced as in the case of $W_{eff}$ (see the
inset of Fig.~\ref{fig:gaussscaling}), the scaling curve coincides
with the non-interacting one, implying there is no $g^{\ast}$ renormalization
from purely gaussian correlations.}

Taken together, these facts imply that there are significant non-gaussian
spatial correlations in the $v_{i}$'s which considerably delay the
crossover to the strongly localized regime. \textcolor{black}{Such
correlations introduce a very exciting new dimension to the physics
of disordered systems, because much of the existing lore about Anderson
localization focused on the effects of random potentials with simple
gaussian statistics - incorrectly assuming that higher-order correlations
play only a secondary role. In the Supplemental Material we further
characterize these inter-site correlations and show how their incorporation
is essential for a $g^{*}\left(U\right)<1$ \cite{suppl}.}

\emph{Extension to higher dimensions.--- }It is tempting to speculate
on what would happen if our main conclusions persist in $d>1$. If
we follow the same phenomenological extension as in Shapiro's work
\cite{PhysRevB.34.4394}, we can write $\tilde{\beta}_{d}\left(g\right)=\beta\left(g\right)+d-1$,
where $\beta\left(g\right)$ is given in Eqs.~\eqref{eq:intbeta}
or \eqref{eq:intbetaand}. Graphically, this corresponds to a vertical
shift of $\beta\left(g\right)$ for $d=2,3$. In particular, for $d=3$,
$\tilde{\beta}_{d}\left(g\right)$ changes sign as expected \cite{Abrahams1979,mackinnon81}.
By construction, $\tilde{\beta}_{d}$ has the correct asymptotic limits:
$\tilde{\beta}_{d}\simeq d-2$ for $g\gg g^{*}$ and $\tilde{\beta}_{d}\propto\ln(g/g^{*})$
for $g\ll g^{*}$. Assuming, as we found, that the main effect of
interactions is to rescale the crossover scale $g^{*}$, the net result
would be to shift this crossover in $\tilde{\beta}_{d}$ to much smaller
conductances. This implies a much more extended ohmic region, even
though at $T=0$ all electronic states should still remain localized
in $d=2$ \cite{fleury08a,fleury08b}. In addition, the proposed interaction-induced
renormalization $g\rightarrow g/g^{*}$ should dramatically reduce
the \emph{amplitude }of the weak-localization correction; precisely
such an effect was observed in $d=2$ magnetoresistance experiments
\cite{kravchdenko2003}. In practice, this would open the possibility
that competing (e.g. Mott or Wigner-Mott) mechanisms for localization
\cite{pankov-2008,camjayi2008} could become dominant well before
Anderson localization effects set in. 

\emph{Conclusions.---} Adding interactions to a disordered system
gives rise to new effects that assist transport even if the single
particle states are all Anderson localized. Our careful numerical
studies show that the typical value of the scaled conductance follows
the same non-interacting behavior but with a large decrease of the
conductance scale $g^{*}\left(U\right)$ signaling the crossover to
the strongly localized regime. Surprisingly, we find that this reduction
is brought about by non-gaussian inter-site correlations, a mechanism
overlooked in previous works. This\textcolor{black}{{} opens an exciting
new door to understanding the effects of interactions in disordered
systems.}

We acknowledge support by DFG through grants FOR 960 and GRK 1621
(ECA), CNPq through grant 304311/2010-3 (EM), FAPESP through grant
07/57630-5 (EM) and NSF through grant DMR-1005751 (VD and HJM). 

\bibliographystyle{apsrev4-1}
\bibliography{all,bibliography}

\end{document}


\title{Supplemental notes for: ``Non-gaussian spatial correlations dramatically
weaken localization''}

\author{H. Javan Mard}

\affiliation{Department of Physics and National High Magnetic Field Laboratory,
Florida State University, Tallahassee, FL 32306}

\author{E. C. Andrade}

\affiliation{Institut für Theoretische Physik, Technische Universität Dresden,
01062 Dresden, Germany}

\author{E. Miranda}

\affiliation{Instituto de Física Gleb Wataghin, Unicamp, R. Sérgio Buarque de
Holanda, 777, Campinas, SP 13083-859, Brazil}

\author{V. Dobrosavljevi\'{c}}

\affiliation{Department of Physics and National High Magnetic Field Laboratory,
Florida State University, Tallahassee, FL 32306}

\date{\today}

\maketitle

\section{Computation of the conductance }

The dimensionless conductance computed in this work is defined as
$g=G/\left(2e^{2}/h\right)$. The sample conductance $g_{s}$, with
the effects of the contacts already removed, is given by the Landauer
relation \cite{markos2006numerical}
\begin{equation}
g_{s}=\frac{<T_{s}>_{typ}}{1-<T_{s}>_{typ}},\label{eq:Landauer}
\end{equation}
where $T_{s}$ is the total sample transmittance, and 'typ' refers
to the geometrical average $<T_{s}>_{typ}=\exp\overline{\ln T_{s}}$
(the overbar denoting an average over realizations). The transmission
function is simply given by \cite{haugjauhobook} 
\begin{eqnarray}
T\left(\omega\right) & = & \frac{1}{2}\mbox{Tr}\left[\mathbf{G}_{S}^{\dagger}\left(\omega\right)\mbox{\boldmath\ensuremath{\Gamma}}_{R}\left(\omega\right)\mathbf{G}_{S}\left(\omega\right)\mbox{\boldmath\ensuremath{\Gamma}}_{L}\left(\omega\right)\right].\label{eq:transmission_function}
\end{eqnarray}
Here, all bold-face quantities are matrices in the lattice site basis,
and $\omega$ is set to zero in our case, representing the Fermi level
$T=0$. $\mbox{\boldmath\ensuremath{\Gamma}}_{L\left(R\right)}\left(\omega\right)$
are the coupling matrices, describing the coupling of the system to
non-interacting left (L) and right (R) leads, and are determined by
the leads' self-energies as 
\begin{eqnarray}
\mbox{\boldmath\ensuremath{\Gamma}}_{L\left(R\right)}\left(\omega\right) & = & i\left(\mbox{\boldmath\ensuremath{\Sigma}}_{L\left(R\right)}-\mbox{\boldmath\ensuremath{\Sigma}}_{L\left(R\right)}^{\dagger}\right).\label{eq:coupling_matrix}
\end{eqnarray}
For the sake of simplicity, we assume 
\begin{eqnarray}
\mbox{\boldmath\ensuremath{\Sigma}}_{L\left(R\right)} & = & -i\eta\mathbf{B}_{L\left(R\right)},\label{eq:constant_self_energy}
\end{eqnarray}
where the matrix $\mathbf{B}_{L\left(R\right)}$ is equal to $1$
for a site $i$ connected to the left (right) lead and $0$ otherwise.
This constant lead self-energy is equivalent to the assumption of
wide lead bands. $\mathbf{G}_{S}$ is the sample Green's function,
given by
\begin{eqnarray}
\mathbf{G}_{S} & = & \left[\left(\omega+\mu\right)\mathbf{1}-\mathbf{H}_{S}^{0}-\mbox{\boldmath\ensuremath{\Sigma}}_{S}-\mbox{\boldmath\ensuremath{\Sigma}}_{L}-\mbox{\boldmath\ensuremath{\Sigma}}_{R}\right]^{-1},\label{eq:G_sample}
\end{eqnarray}
where $\mu$ is the chemical potential. $\mathbf{H}_{S}^{0}$ is the
non-interacting sample Hamiltonian and $\mbox{\boldmath\ensuremath{\Sigma}}_{S}$
is the sample self-energy, which accounts for the electronic correlations.
As we employ a Hartree-Fock approach, the sample self-energy is given
by
\begin{eqnarray}
\mbox{\boldmath\ensuremath{\Sigma}}_{S} & = & \frac{U}{2}\left\langle \mathbf{n}\right\rangle +\mu\mathbf{1}.\label{eq:self_energy_SB4}
\end{eqnarray}
$\mathbf{n}$ is a diagonal matrix in the site basis whose elements
$n_{i}$ correspond to the occupation of the site $i$. It can be
seen from Eqs.~(\ref{eq:G_sample}), (\ref{eq:self_energy_SB4})
and (\ref{eq:constant_self_energy}) that the sample Green's function
can be obtained through a single diagonalization of an effective Hamiltonian,
as opposed to a matrix inversion for every frequency value. Even with
this simplification, we are still left with a non-Hermitian matrix.
We have implemented this diagonalization step using standard LAPACK
routines. A particular efficient implementation can be achieved using
the OpenMP version of Intel\textquoteright s MKL library, which allowed
us to perform these calculations on a desktop computer. Furthermore,
using Eqs.~(\ref{eq:coupling_matrix}) and (\ref{eq:constant_self_energy}),
we can rewrite Eq.~(\ref{eq:transmission_function}) as
\begin{eqnarray}
T\left(\omega\right) & = & 2\eta^{2}\mbox{Tr}\left[\mathbf{G}_{1L}\left(\omega\right)\mathbf{G}_{L1}^{\dagger}\left(\omega\right)\right],\label{eq:transmission_simplified}
\end{eqnarray}
where $\mathbf{G}_{1L}\left(\omega\right)$ is the sample Green's
function from the first site ($i=1$) (connected to the left lead)
to the last one ($i=L$) (connected to the right lead).

\section{\textcolor{black}{The localization length }}

\begin{figure}[H]
\includegraphics[width=3.5in]{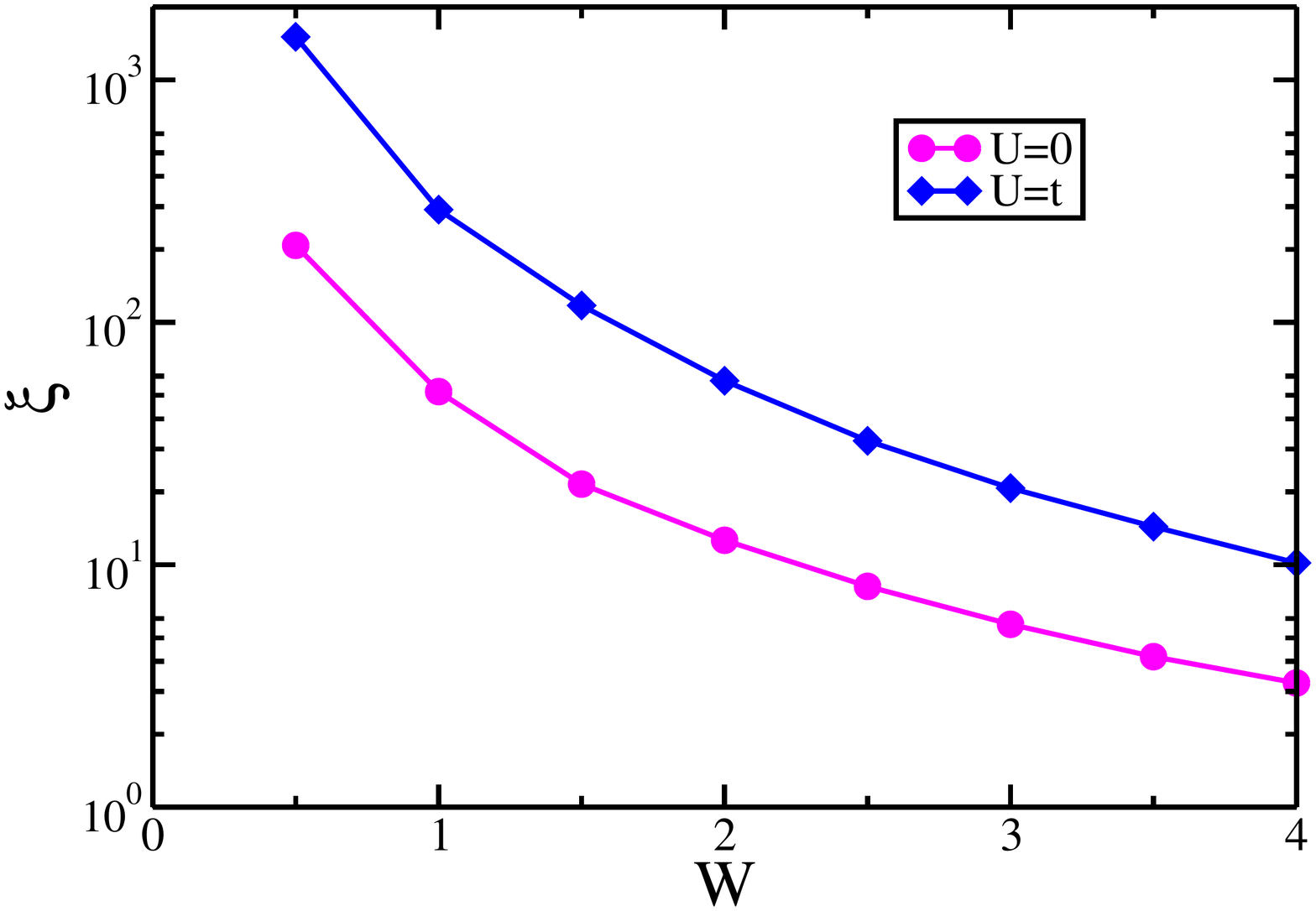}

\protect\caption{\label{fig:xiplotu1} (Color online) Localization length $\xi$ as
a function of $W$ obtained from the collapse of the conductance curves.
$\xi$ is strongly enhanced for $U=t$, even though all states remain
localized.}
\end{figure}

\textcolor{black}{The localization length $\xi$ is calculated using
the exponential behavior of the conductance $(e^{-x})$ at very large
$x=\frac{L}{\xi}$ (for this, either $L$ or $W$ should be very large).
In other words, it can be computed from the slope of the conductance
as a function of $x$ on a semi-log scale (assuming all states are
localized, which we always find). Fig.~\ref{fig:xiplotu1} shows
that adding interactions considerably increases the localization length,
even though all states remain localized.}

\section{Operational definition of the characteristic conductance}

We now show that the characteristic conductance $g^{*}$ can be determined
in practice as the value at the point $L=L^{*}=(\ln2)\xi$ , or $x=\ln2$.
Given the beta function for 1D localization, we can write

\begin{equation}
\ln\frac{L}{L_{0}}=\int_{g_{0}}^{g}\frac{d\ln g}{\beta\left(g\right)}=F\left(g\right)-F\left(g_{0}\right),\label{eq:lnL}
\end{equation}
where we define the function
\begin{equation}
F\left(g\right)\equiv\int^{g}\frac{d\ln g}{\beta\left(g\right)},\label{eq:F(g)def}
\end{equation}
and $g_{0}$ is a reference conductance at scale $L_{0}$. From Eq.
(\ref{eq:lnL}) we get 

\begin{equation}
F\left[g\left(L\right)\right]=\ln\left[\frac{Le^{F\left(g_{0}\right)}}{L_{0}}\right],\label{eq:F(g)simplified}
\end{equation}
which implicitly gives $g\left(L\right)$. Defining the reference
conductance $g=g^{*}$ at $L=L^{*}$ 

\begin{equation}
F\left[g\left(L\right)\right]=\ln\left(\frac{Le{}^{F\left(g^{\ast}\right)}}{L^{*}}\right).\label{eq:F(g)definedbyg*}
\end{equation}
Using the form proposed in the main text $\beta\left(g\right)=\beta_{0}\left(g/g^{\ast}\right)$
and the non-interacting beta function of ref.~\onlinecite{Anderson1980},
we find 

\begin{equation}
F\left(g\right)=-\int^{g}\frac{d\ln g}{\left(1+\frac{g}{g^{*}}\right)\ln\left(1+\frac{g^{*}}{g}\right)}=\ln\left[B\ln\left(1+\frac{g^{*}}{g}\right)\right],\label{eq:F(g)Andersonf}
\end{equation}
where $B$ is an arbitrary constant. It then follows from Eq.~(\ref{eq:F(g)definedbyg*})
that

\begin{equation}
F(g)=\ln(\frac{L}{L^{*}}B\ln2).\label{eq:F(g)at g*}
\end{equation}
On the other hand, the scaling function proposed in ref.~\onlinecite{Anderson1980}
is

\begin{equation}
\frac{g^{*}}{g}=e^{L/\xi}-1,\label{eq:gformulag*}
\end{equation}
which, when plugged into Eq.~(\ref{eq:F(g)Andersonf}), leads to

\begin{equation}
F(g)=\ln[B\frac{L}{\xi}].\label{eq:F(g)intract}
\end{equation}
Finally, comparing Eqs.~(\ref{eq:F(g)at g*}) and (\ref{eq:F(g)intract})
gives us the anticipated result $L^{*}=(\ln2)\xi\thickapprox0.69\xi.$

\section{Spatial correlations of the renormalized disorder potential}

\subsection{Weak disorder}

An important question is how to characterize the spatial correlations
between the renormalized site energies $v_{i}$. For weak disorder,
this problem is solvable analytically. Expanding the average on the
right-hand side of Eq.~(2) of the main text to first order in the
bare disorder, we find the spatial Fourier component \cite{eric10,eric12}

\begin{equation}
v_{q}=\frac{\varepsilon_{q}}{1-U\Pi_{q}}+O(\varepsilon_{q}^{2}),\label{eq:vqlinearexp}
\end{equation}
where $\Pi_{q}$ is the usual static Lindhard polarization function
of the clean, non-interacting system, which in the case of a one-dimensional
tight-binding chain is given by 
\begin{eqnarray}
\Pi_{q} & = & -\rho\left(\varepsilon_{F}\right)\frac{1}{\mbox{sin}\left(q/2\right)}\mbox{ln}\left[\frac{\mbox{cos}\left(q/2\right)}{1-\mbox{sin}\left(q/2\right)}\right],\label{eq:lind1d}
\end{eqnarray}
where $\rho\left(\varepsilon_{F}\right)$ is the density of states
at the Fermi level.

\begin{figure}[h]
\begin{centering}
\includegraphics[scale=0.3]{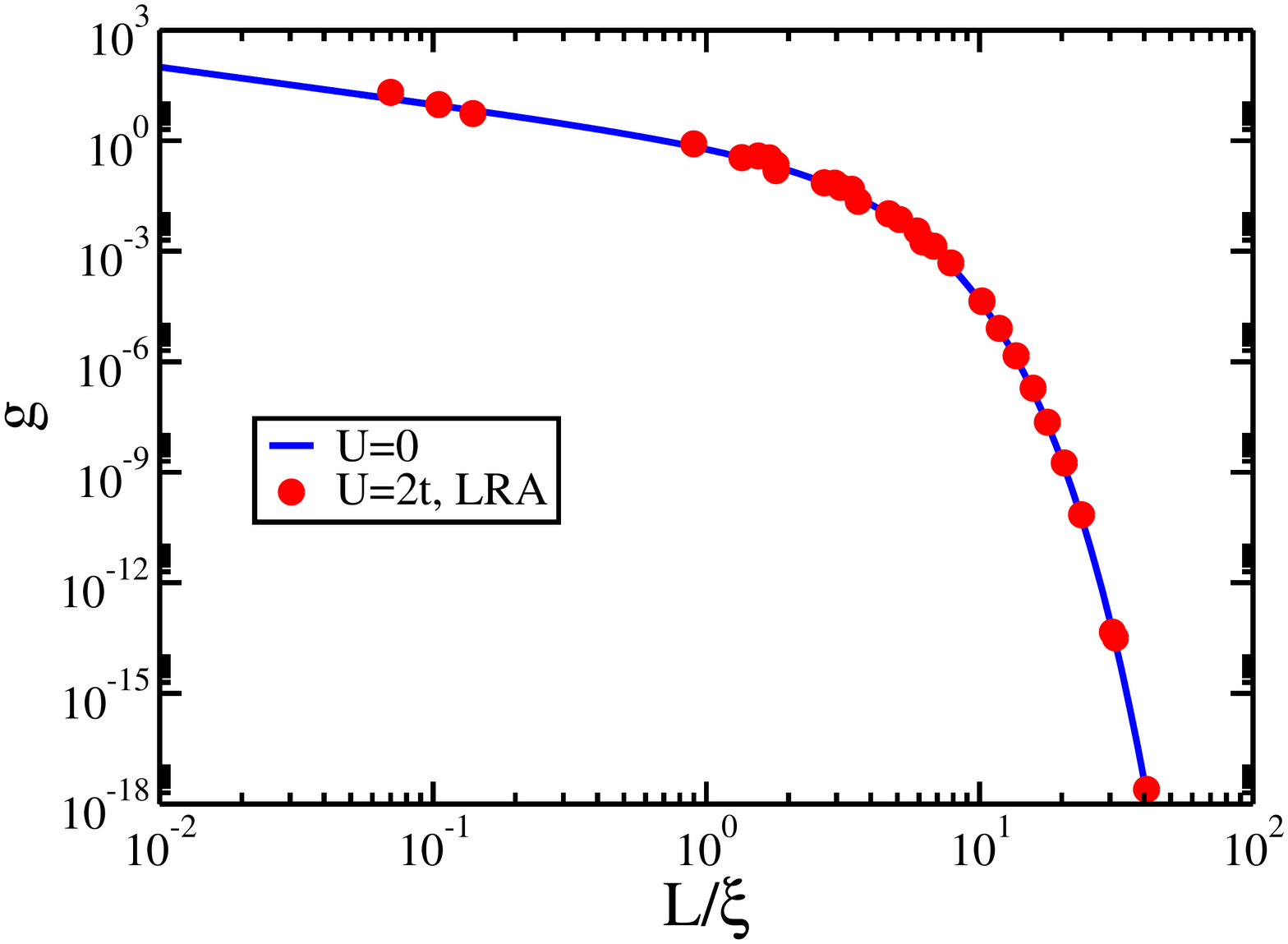}
\par\end{centering}

\protect\caption{\label{fig:LRAplot} (Color online) The scaling function in the linear
response approach compared with the non-interacting one. }
\end{figure}

Using Eqs.~(\ref{eq:vqlinearexp}) and (\ref{eq:lind1d}) we can
generate a renormalized potential $v_{i}$ in this linear response
approach (LRA). The result of this procedure is shown in Fig.~\ref{fig:LRAplot}.
As can be seen from Fig.~\ref{fig:LRAplot}, although the LRA potential
exhibits both disorder screening and spatial correlations, it does
not capture the renormalization of $g^{*}$: \textcolor{black}{the
scaling function (and its mathematical equivalent, the beta function)}\textcolor{blue}{{}
}coincides with the one from the non-interacting calculation. This
result strongly suggests that the $g^{*}$ renormalization is a \emph{non-perturbative}
effect of disorder and a fully self-consistent solution is needed
in order to capture it.

\subsection{Gaussian correlations}

\begin{figure}[h]
\begin{centering}
\includegraphics[width=3.5in]{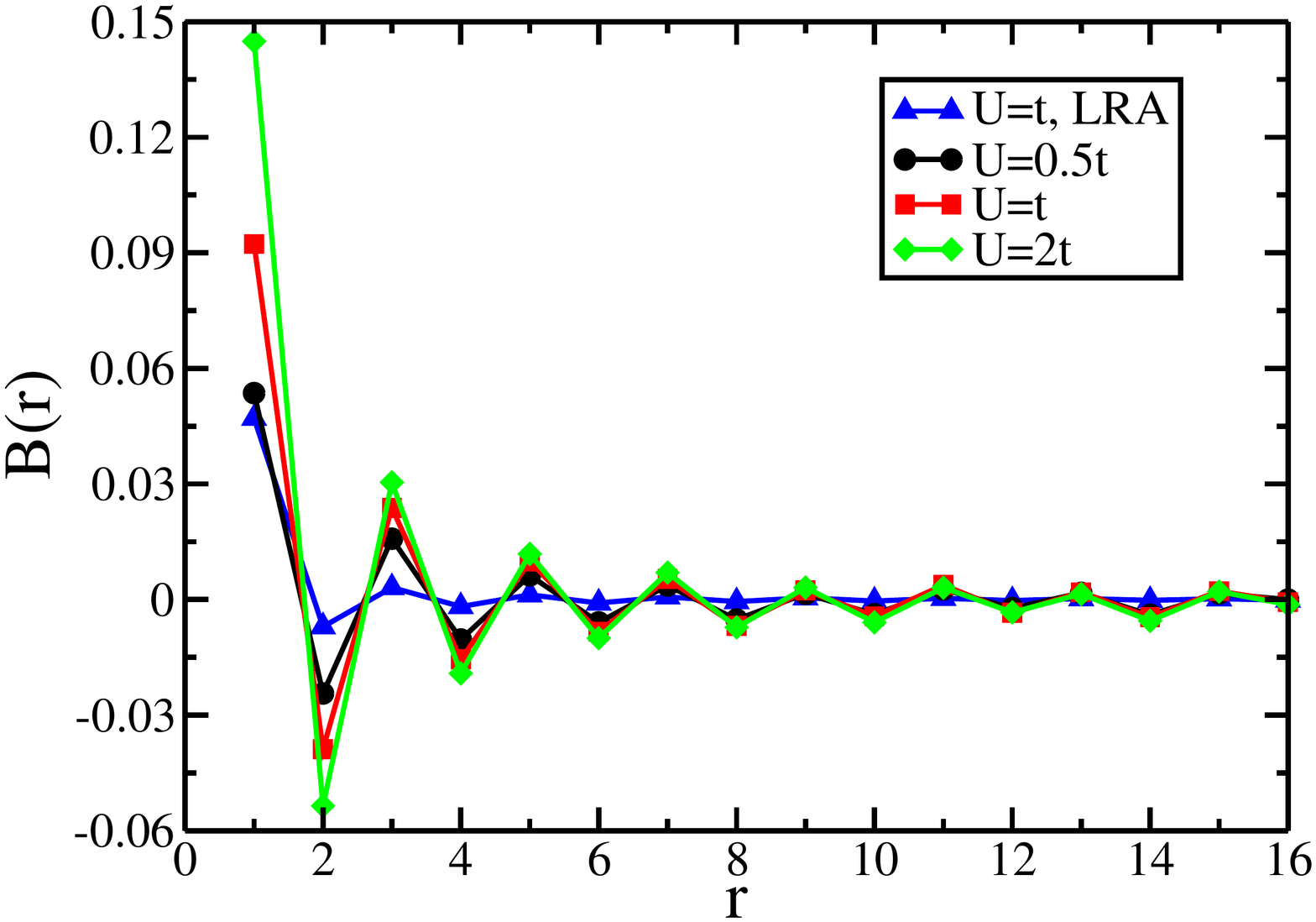}
\par\end{centering}

\protect\caption{\label{fig:Friedel }(Color online) Spatial pair correlations between
the renormalized site energies for $W=1.0t$ and $L=400$.}
\end{figure}

To go beyond the weak disorder limit, we calculate \textcolor{black}{the
fully self-consistent HF two-point correlation function $\left\langle v_{i}v_{j}\right\rangle $.
Here $\left\langle \cdots\right\rangle $} denotes an average over
both pairs and disorder realizations. This function only depends on
the distance $r=r_{i}-r_{j}$ between the two sites. In Fig.~\ref{fig:Friedel }
we show $B\left(r\right)=\left\langle v_{i}v_{j}\right\rangle /W_{eff}^{2}$
as a function of the distance $r$ between two sites. $W_{eff}$ is
the effective disorder width, which is simply given by the standard
deviation of $v_{i}$. Defined in this fashion, $B\left(r=0\right)=1$.
It displays Friedel-like oscillations which are enhanced as $U$ increases.\textcolor{magenta}{{}
}\textcolor{black}{For comparison, we also show $B\left(r\right)$
obtained within LRA at $U=t$.}

\begin{figure}[H]
\begin{centering}
\includegraphics[scale=0.3]{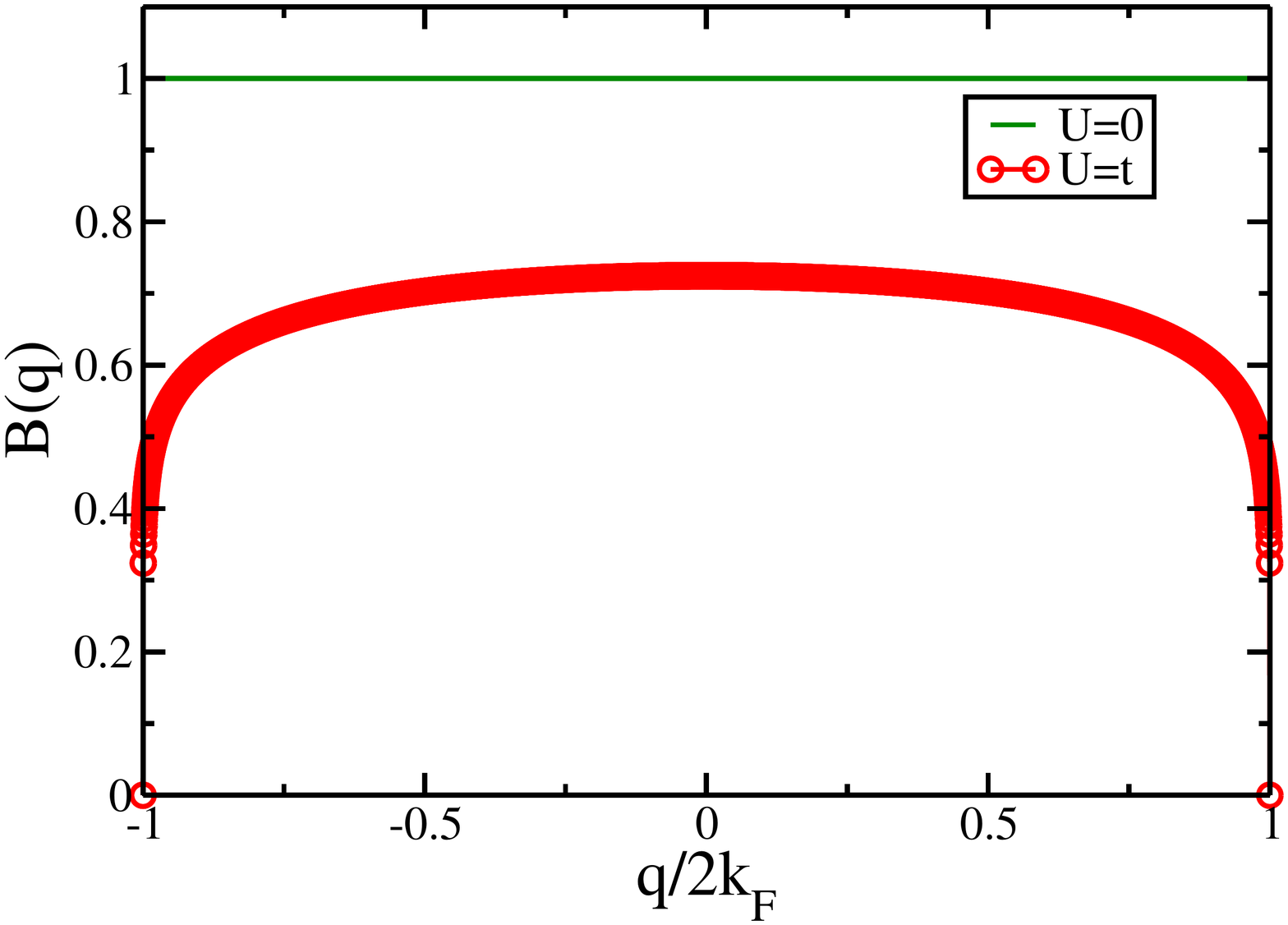}
\par\end{centering}

\protect\caption{\label{fig:Bq} (Color online) Fourier transform of the two point
correlation function $B_{ij}$ for $W=2t$ and $U=t$.}
\end{figure}

Since a Gaussian distribution is fully determined by its two-point
correlation function, we can use \textcolor{black}{$\left\langle v_{i}v_{j}\right\rangle $}
to generate Gaussian correlated variables. More precisely, we generated
$v_{i}$ deviates obeying a multivariate Gaussian distribution with
zero mean and covariance matrix \textcolor{black}{$\left\langle v_{i}v_{j}\right\rangle $.
}To do this, we first calculate the Fourier transform of $B_{ij}$,
$B(q)$, which is shown in Fig.~\ref{fig:Bq}. As expected, its derivative
diverges at $q=\pm2k_{F}$, giving rise to long-ranged Friedel-like
oscillations, as shown in Fig.~\ref{fig:Friedel }. \textcolor{black}{Once
we have $B(q)$, we generate Gaussian distributed random complex numbers
$v_{q}$ using the Box-Muller method with zero mean and a $q$-dependent
variance $W_{eff}^{2}B(q)$ \cite{Box-Muller}. Note that each $v_{q}$
obeys a univariate Gaussian distribution. We can then obtain their
real space values $v_{i}$ after a numerical Fourier transform.}\textcolor{blue}{{} }

\begin{figure}[b]
\includegraphics[width=3.5in]{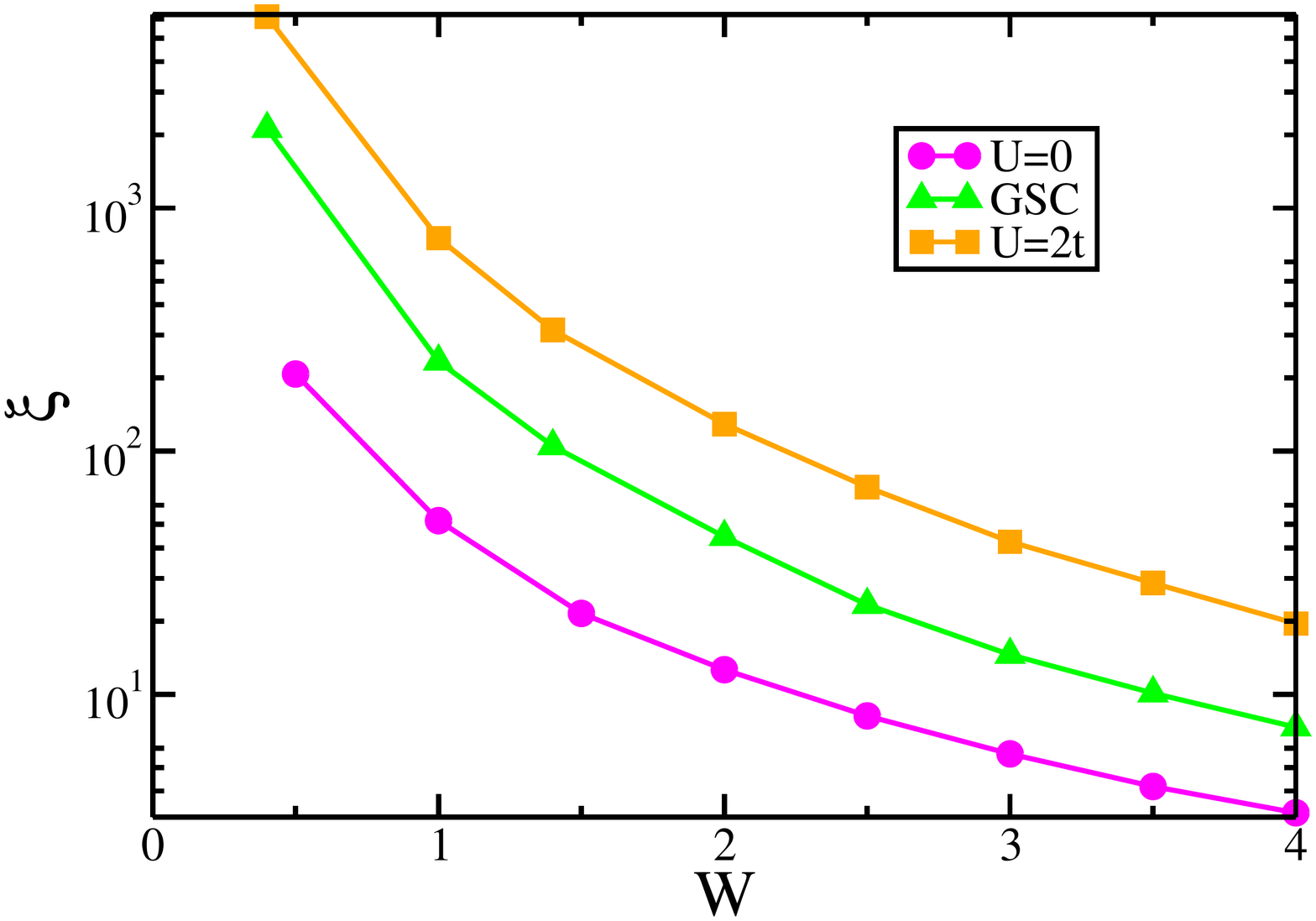}

\protect\caption{\label{fig:xiplotgGSC}(Color online) Localization length $\xi$ as
a function of $W$, obtained from the collapse of the conductance
curves. We see that $\xi$ is enhanced if we retain only Gaussian
correlations of the site energies (GSC), but not as strongly as to
be able to reproduce the complete HF results.}

\end{figure}

The result of this procedure is shown in the main text, where it is
shown to be insufficient to capture both the renormalization of the
crossover scale $g^{*}$ and the enhancement of the localization length
(see Fig.~\ref{fig:xiplotgGSC}). Therefore, we conclude that this
phenomenon is not only non-perturbative in disorder but also determined
by higher order disorder correlations beyond the Gaussian level.

\subsection{Full characterization of inter-site correlations}

\begin{figure}[t]
\includegraphics[scale=0.5]{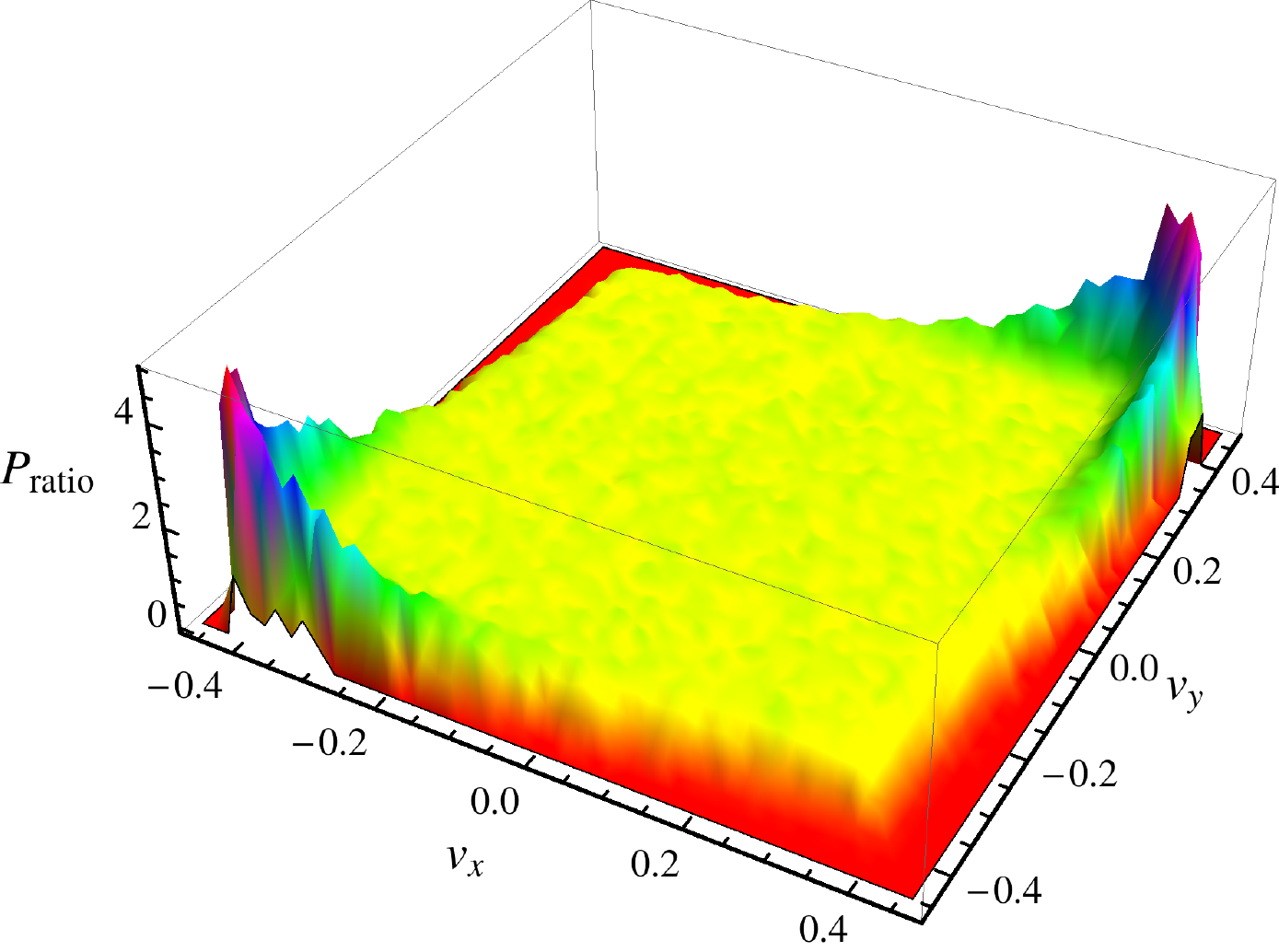}

\protect\caption{\label{fig:p2overpipi} (Color online) The ratio of distributions
$P_{ratio}(x,y)=P_{1}^{(2)}\left(x,y\right)/P^{\left(1\right)}\left(x\right)P^{\left(1\right)}\left(y\right)$
for $U=W=t$, and $L=10^{3}$ with $1,000$ realizations of disorder.
Deviations from 1 are indications of inter-site correlations.}
\end{figure}

In order to characterize the behavior of the inter-site correlations,
we focus on the distribution function of two given sites $i$ and
$j$. Evidently, this distribution only depends on the distance between
sites $\left|i-j\right|=r$, so we will denote it by $P_{r}^{\left(2\right)}\left(v_{i}\equiv x,v_{j}\equiv y\right)$.
If there were no inter-site correlations, then we would have 
\begin{equation}
P_{r}^{\left(2\right)}\left(x,y\right)=P^{\left(1\right)}\left(x\right)P^{\left(1\right)}\left(y\right),\label{eq:uncorrp2}
\end{equation}
where $P^{\left(1\right)}\left(v_{i}\equiv x\right)$ is the distribution
of renormalized site energies of a given site. Both distributions
were obtained numerically from our calculations. In Fig.~\ref{fig:p2overpipi}
we show the ratio of the left-hand and the right-hand sides of Eq.~(\ref{eq:uncorrp2})
for $r=1$. The fact that it is not equal to 1 is a demonstration
of the existence of inter-site correlations.

Further insight can be gained by noting that if Eq.~(\ref{eq:uncorrp2})
were true, then $P_{r}^{\left(2\right)}\left(x,y\right)$ would symmetric
with respect to the interchange of its arguments. That this is not
the case is made clear by a glance at a color scale plot of $P_{r}^{\left(2\right)}\left(x,y\right)$
in the $xy$ plane, as shown in Fig.~\ref{fig:p(vi,vj)} for $r=1$.
In fact, this anisotropy suggests a simple parametrization of the
distributions.

\begin{figure}[t]
\includegraphics[scale=0.8]{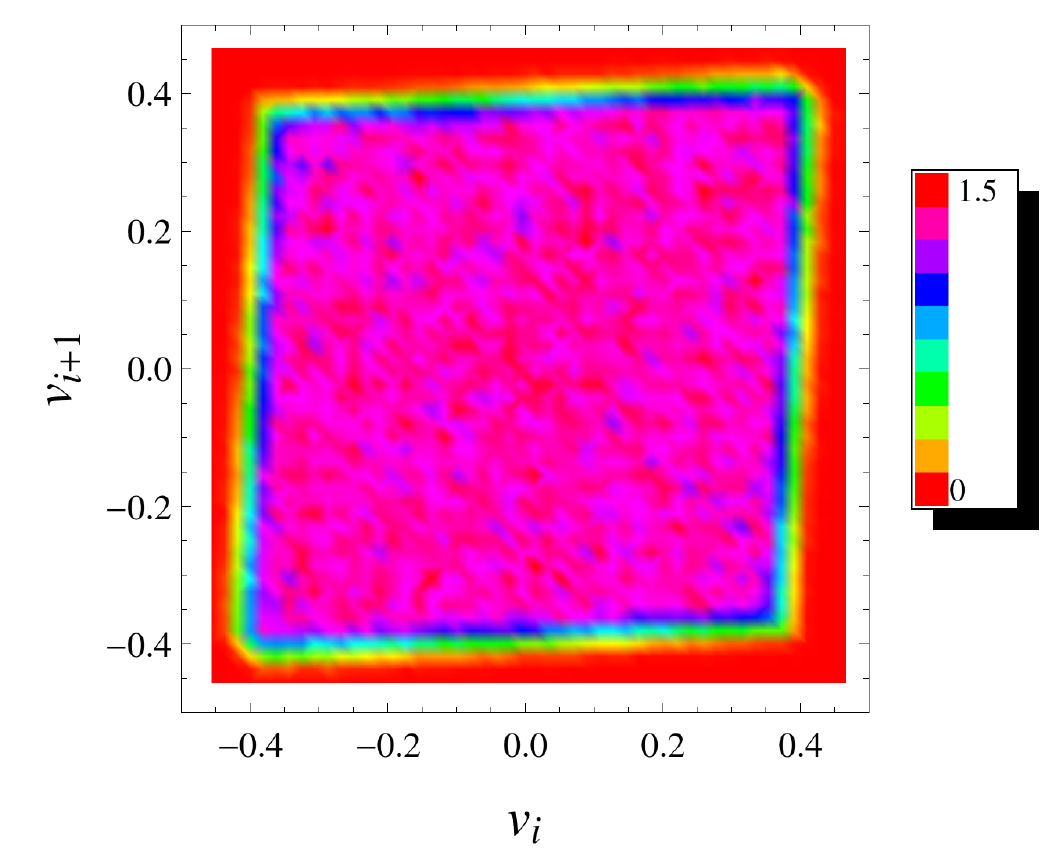}

\protect\caption{\label{fig:p(vi,vj)} (Color online) The color scale plot of $P_{1}^{\left(2\right)}\left(x,y\right)$
for $U=W=t$, and $L=10^{3}$ with $1,000$ realizations of disorder.
The strong anisotropy reflects the presence of appreciable inter-site
correlations.}
\end{figure}

\begin{figure}[t]
\begin{centering}
\includegraphics[width=3.5in]{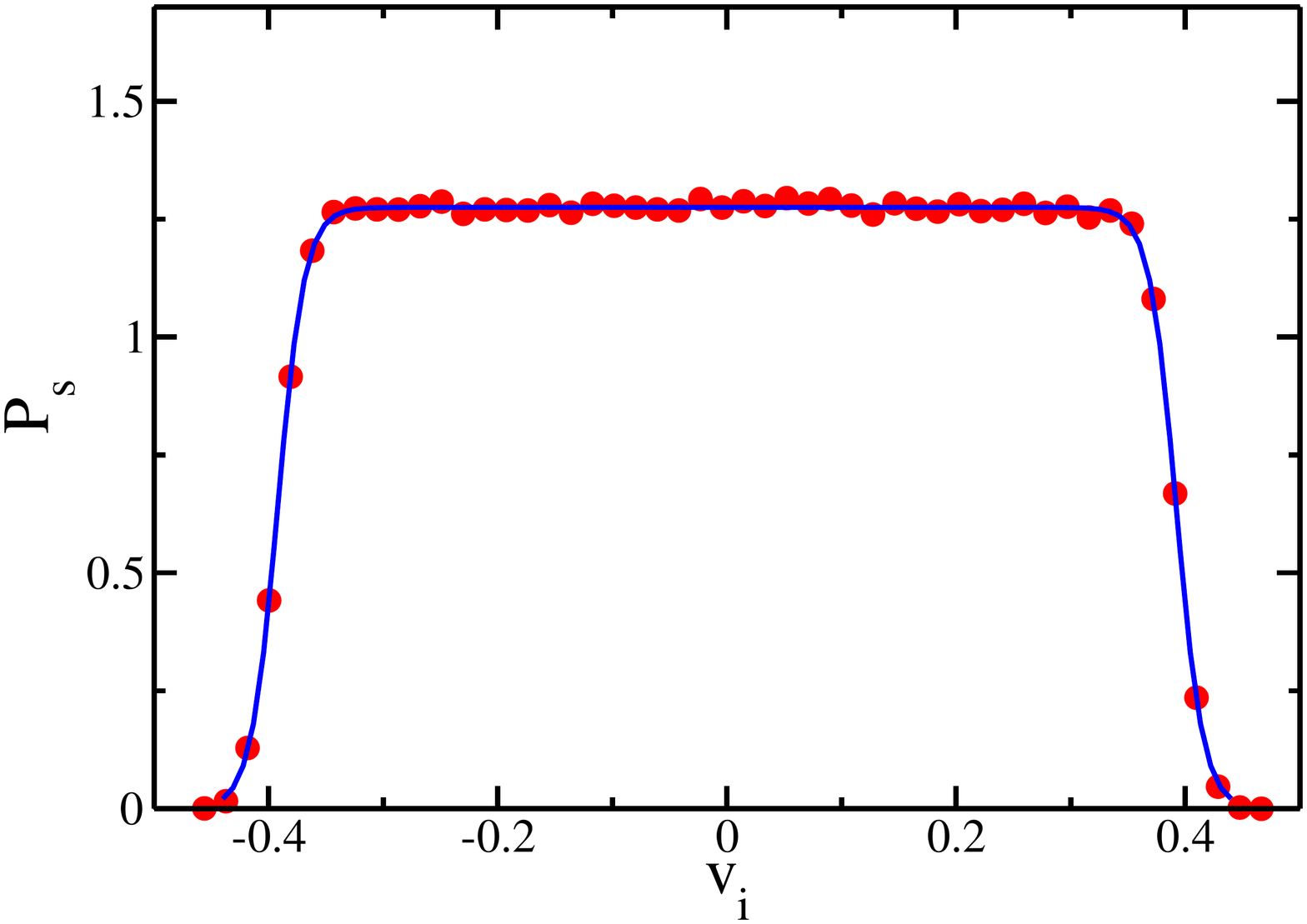}
\par\end{centering}

\protect\caption{\label{fig:P(vi)} (Color online) The histogram of single site energy,
$P\left(v_{i}\right)$, is well fitted to a Fermi-Dirac like distribution.
The solid line is the fitted function described by Eq. \ref{eq:fermi-diracformula}
with two parameters $\widetilde{W}=0.3921\pm0.0002$, and $\delta\widetilde{W}=0.01172\pm0.0001$.
Here $U=W=t$, and $L=1000$.}
\end{figure}

We first notice that the $P^{\left(1\right)}\left(x\right)$ can be
very accurately captured by the following two-parameter function if
the bare disorder is uniform 
\begin{equation}
Q^{\left(1\right)}(x)=\frac{1/\widetilde{W}}{1+\mbox{exp}\left[\left(|x|-\widetilde{W}/2\right)/\delta\widetilde{W}\right]}.\label{eq:fermi-diracformula}
\end{equation}
Here, $\widetilde{W}$ is the effective renormalized disorder and
$\delta\widetilde{W}$ rounds the tails of the distribution. Fig.~\ref{fig:P(vi)}
shows the raw numerical data compared to the best fit using Eq.~(\ref{eq:fermi-diracformula}).
The agreement is excellent. 

We then propose a single-parameter function for the two-site distribution
\begin{equation}
Q_{r}^{\left(2\right)}\left(x,y\right)=Q^{\text{\ensuremath{\left(1\right)}}}\left(x^{\prime}\right)Q^{\text{\ensuremath{\left(1\right)}}}\left(y^{\prime}\right),\label{eq:2parQ2}
\end{equation}
where 
\begin{equation}
\begin{cases}
x^{\prime} & =\frac{1}{1-c^{2}}x-\frac{c}{1-c^{2}}y,\\
y^{\prime} & =\frac{-c}{1-c^{2}}x+\frac{1}{1-c^{2}}y,
\end{cases}\label{eq:transformation}
\end{equation}
and $c$ is the anisotropy parameter. The above transformation represents
a stretch along the main diagonal and a shrinking along the secondary
diagonal. An excellent description of the data is obtained with Eq.~(\ref{eq:2parQ2}).

\begin{figure}
\includegraphics[scale=0.15]{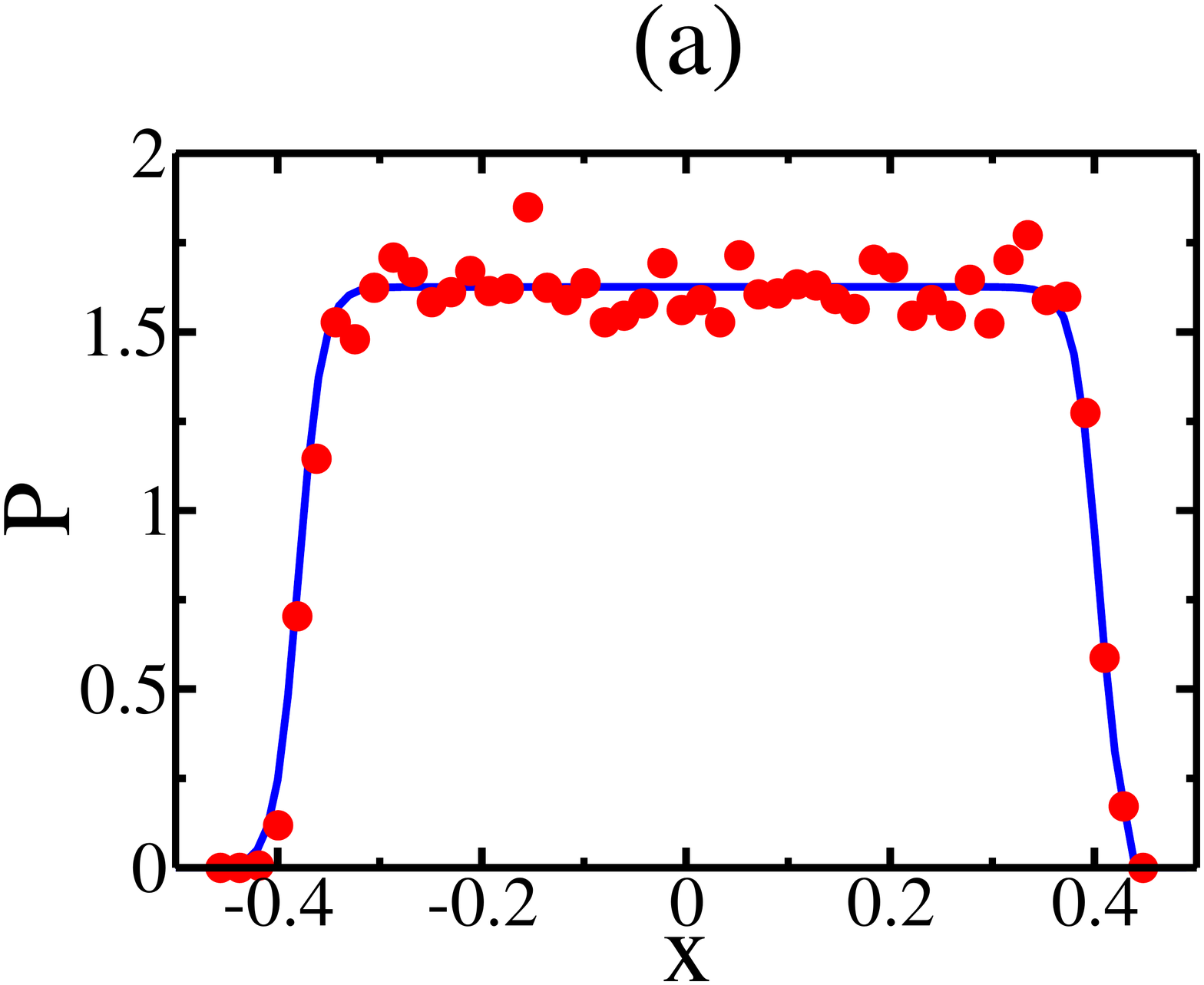} \includegraphics[scale=0.15]{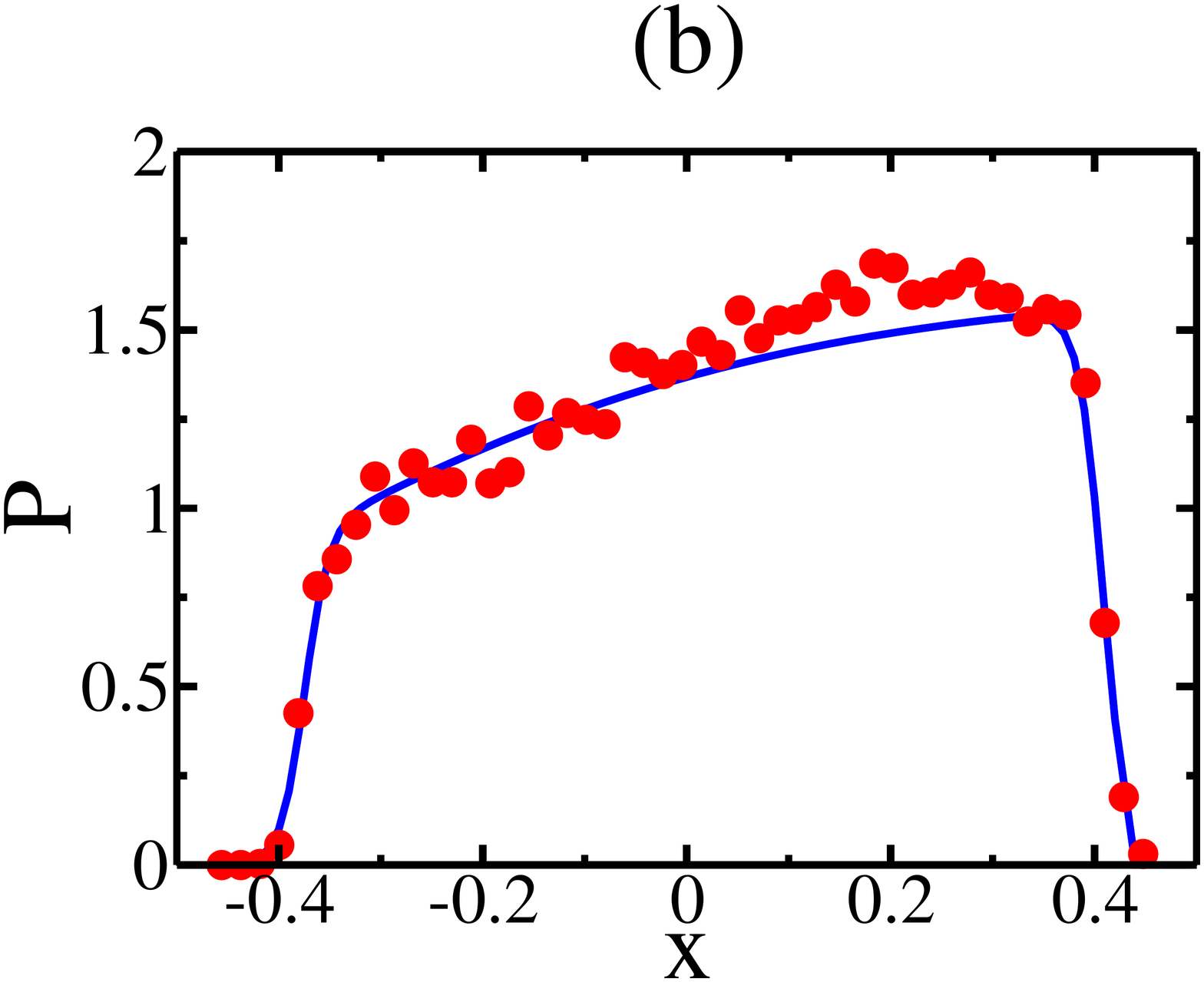}

\protect\caption{The proposed function $Q_{r=1}^{\left(2\right)}\left(x,y\right)$
(blue lines) compared to the raw data (red dots) for fixed values
of $y$: (a) $y=0.27$ and (b) $y=0.37$. }

\end{figure}

\begin{figure}[t]
\includegraphics[scale=0.32]{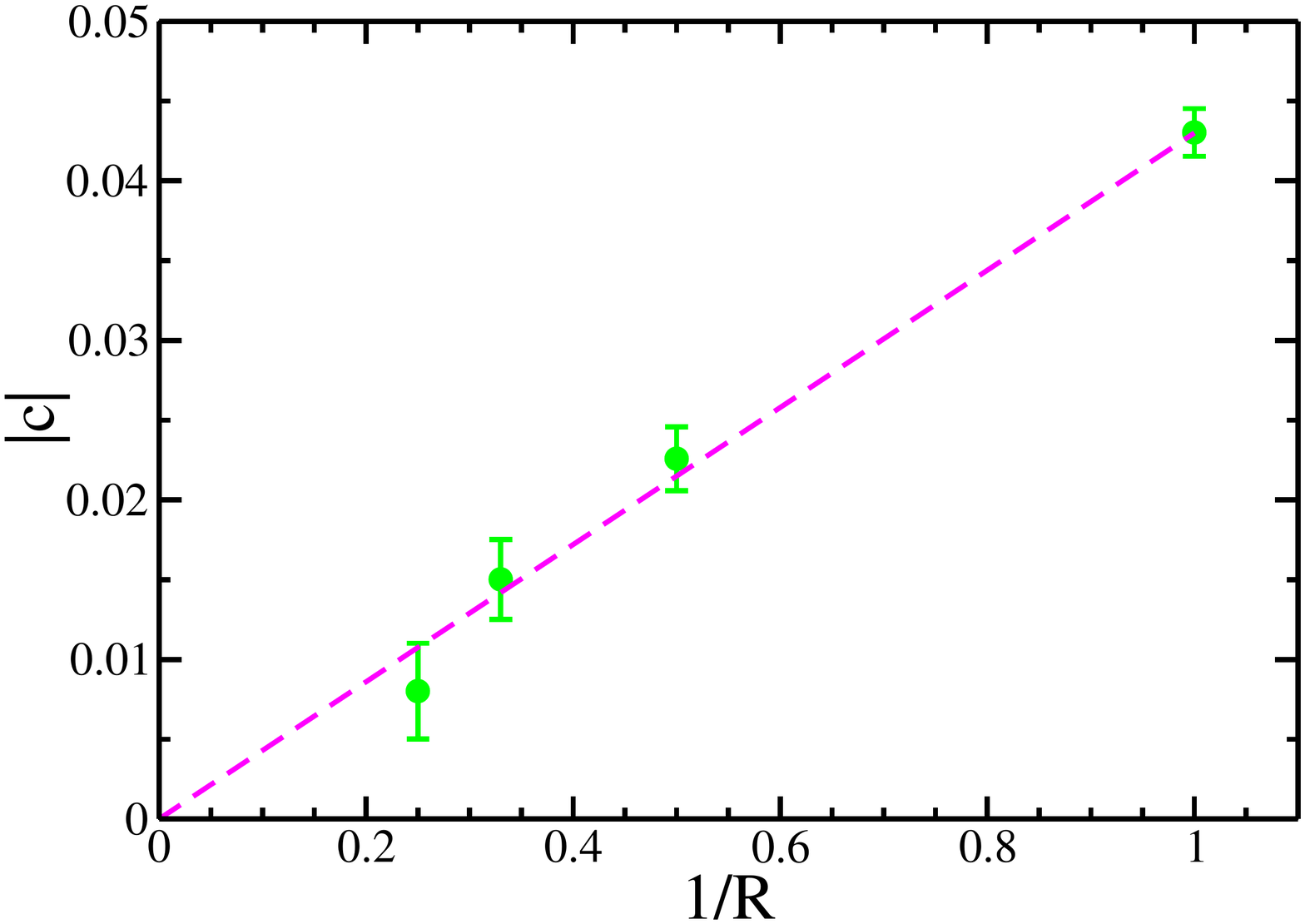}

\protect\caption{\label{fig:c-fitting} (Color online) The anisotropy parameter $c$
as a function of the inter-site distance. The dashed line fit is $|c|=\alpha/R$,
where $\alpha=0.045\pm0.003$. Here $U=W=t$, and $L=1000$.}
\end{figure}

Therefore, we can parametrize the two-site correlations by the single
parameter $c$. The dependence of $c$ on $r$ is shown in Fig.~\ref{fig:c-fitting},
where we see a gradual decrease of correlations with the distance
between the two sites. This reduction is well fitted by a straight
line and our results are consistent with $c\rightarrow0$ as $r\rightarrow\infty$.

Since the most important inter-site correlations are for nearest-neighbor
sites, we have generated random site energies distributed according
to $P_{1}^{\left(2\right)}\left(x,y\right)$, thus neglecting correlations
for $r>1$. We then calculated the corresponding conductance, which
we call $g_{nn}$, `nn' here highlighting the fact that it contains
the effects of the \emph{exact} correlations up to nearest neighbors. 

\begin{figure}[t]
\begin{centering}
\includegraphics[scale=0.32]{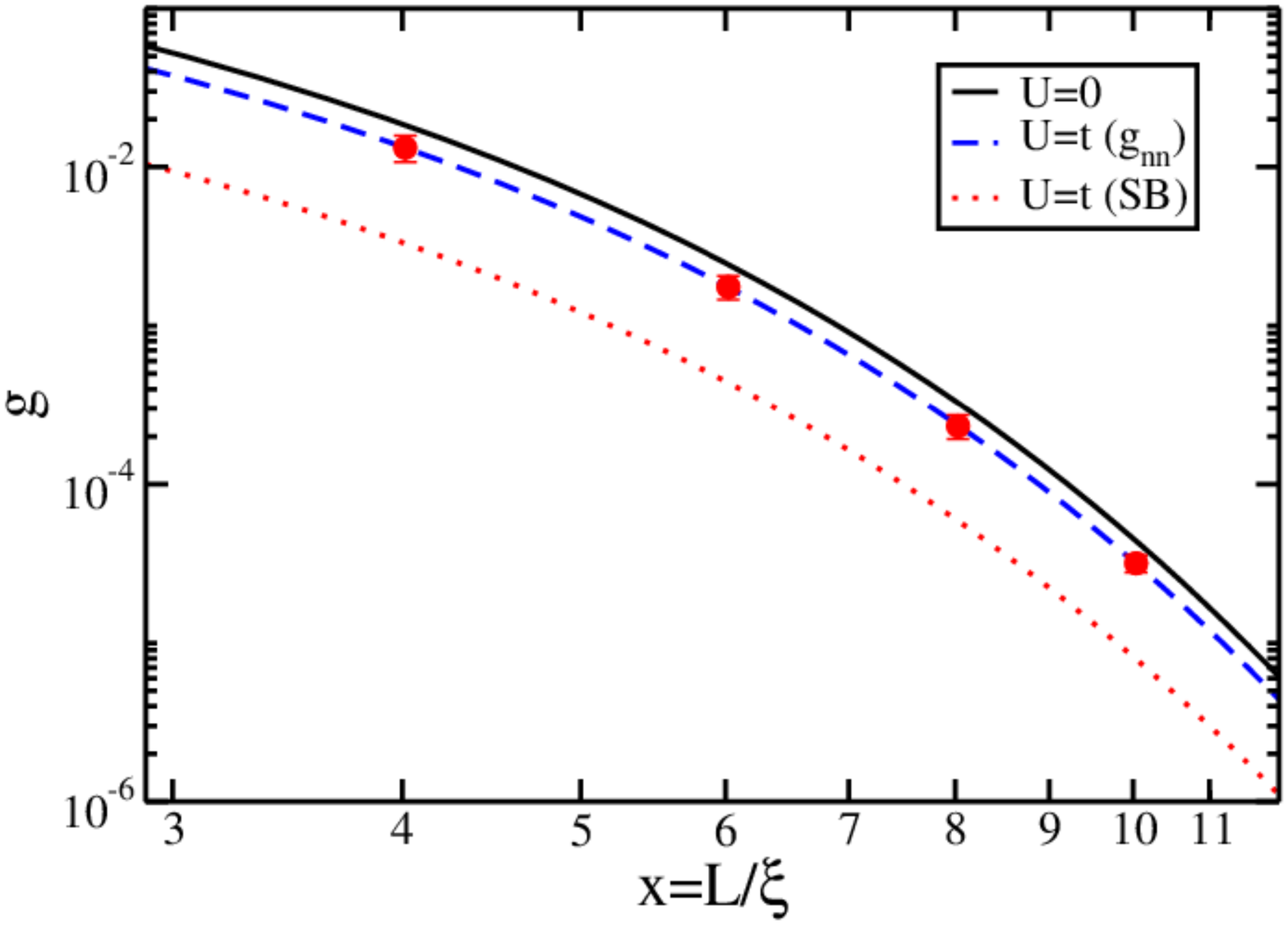}
\par\end{centering}

\protect\caption{\label{fig:g_nn scaling} (Color online) The scaling plot of $g_{nn}$
(see text) as a function of $x=L/\xi$ for $U=t$ (red circles), fitted
to the scaling function of Eq.~(\ref{eq:gformulag*}) (dashed blue
line), with characteristic conductance $g^{*}=0.72$. Also shown are
the scaling functions for the non-interacting case ($g^{*}=0$, continuous
black line) and for the full slave boson results at $U=t$ ($g^{*}=0.18$,
dotted red line). The localization length $\xi$ was obtained form
the exponential behavior at large system sizes ($g\propto\exp(-L/\xi)$).}
\end{figure}

The scaling plot of $g_{nn}$ is shown as the red circles in Fig.~\ref{fig:g_nn scaling}.
We have fitted the data points to the rescaled conductance of Eq.~(\ref{eq:gformulag*})
(dashed blue line). For comparison, we have also plotted the corresponding
scaling curves of the full (slave boson) calculation (dotted red line)
and of the non-interacting case (continuous black line). The characteristic
conductance of $g_{nn}$ is $g^{*}=0.72$, which is significantly
different from the non-interacting value of 1, even though it is still
larger than the full value of $0.18$. Clearly, inter-site correlations
are responsible for the suppression of $g^{*}$. Naturally, spatial
correlations with $r>1$ should to be taken into account to recover
the full value of $g^{*}$. 

As advertised at the beginning of this section, our results show that
the spatial correlations among the renormalized site energies $v_{i}$
are large for $r=1$. Novel effects coming from a short-ranged form
of inter-site correlations of the screened disorder potential have
been discussed in the literature before, for instance in the context
of the random dimer model \cite{dunlap90}, where the presence of
extended states in $1D$ is firmly established \cite{dunlap90}. Even
though our model does not show extended states, the renormalization
of $g^{*}$ translates into a delay in the crossover from extended
to the localized states, strongly hinting towards a link between inter-site
correlations and the robustness of extended states in disordered electronic
systems. 

\bibliographystyle{apsrev4-1}
\bibliography{bibliography,all}